\documentclass{emulateapj}
\usepackage{apjfonts,pstricks}

\newcommand{\msun}{\mbox{ M$_{\odot}$}}

\newcommand{\photflux}{\mbox{ photons cm$^{-2}$ s$^{-1}$ sr$^{-1}$}}

\newcommand{\photfluxA}{\mbox{ photons cm$^{-2}$ s$^{-1}$ sr$^{-1}$
    \AA$^{-1}$}}
\newcommand{\cmden}{\mbox{ cm$^{-3}$}}
\newcommand{\colden}{\mbox{ cm$^{-2}$}}
\newcommand{\kms}{\mbox{ km s$^{-1}$}}
\newcommand{\secinv}{\mbox{ s$^{-1}$}}
\newcommand{\Mpc}{\mbox{ Mpc}}
\newcommand{\kpc}{\mbox{ kpc}}
\newcommand{\keV}{\mbox{ keV}}
\newcommand{\hunits}{\mbox{ km s$^{-1}$ Mpc$^{-1}$}}
\newcommand{\kel}{\mbox{ K}}
\newcommand{\osix}{\ion{O}{6} $\lambda 1032$ }
\newcommand{\osixper}{\ion{O}{6} $\lambda 1032$}
\newcommand{\cfour}{\ion{C}{4} $\lambda 1548$ }
\newcommand{\cfourper}{\ion{C}{4} $\lambda 1548$}
\newcommand{\bq}{\begin{equation}}
\newcommand{\eq}{\end{equation}}

\begin{document}

\title{Ultraviolet Line Emission from Metals in the Low-Redshift
  Intergalactic Medium}

\lefthead{METAL LINES IN THE IGM}
\righthead{FURLANETTO ET AL.}

\author{Steven R. Furlanetto\altaffilmark{1}, Joop
Schaye\altaffilmark{2}, Volker Springel\altaffilmark{3}, \& Lars
Hernquist\altaffilmark{4}}

\altaffiltext{1} {Division of Physics, Mathematics, \& Astronomy;
  California Institute of Technology; Mail Code 130-33; Pasadena, CA
  91125; sfurlane@tapir.caltech.edu}

\altaffiltext{2} {School of Natural Sciences, Institute for Advanced
Study, Einstein Drive, Princeton NJ 08540; schaye@ias.edu}

\altaffiltext{3} {Max-Planck-Institut f\"{u}r Astrophysik,
Karl-Schwarzschild-Strasse 1, 85740 Garching, Germany;
volker@mpa-garching.mpg.de }

\altaffiltext{4} {Harvard-Smithsonian Center for Astrophysics, 60
Garden St., Cambridge, MA 02138; \\ lhernquist@cfa.harvard.edu }

\begin{abstract}
  
  We use a high-resolution cosmological simulation that includes
  hydrodynamics, multiphase star formation, and galactic winds to
  predict the distribution of metal line emission at $z \sim 0$ from
  the intergalactic medium (IGM).  We focus on two ultraviolet doublet
  transitions, \ion{O}{6} $\lambda \lambda 1032,1038$ and \ion{C}{4}
  $\lambda \lambda 1548,1551$.  Emission from filaments with moderate
  overdensities is orders of magnitude smaller than the background,
  but isolated emission from enriched, dense regions with $T \sim
  10^5$--$10^{5.5} \kel$ and characteristic size $\sim 50$--$100 \kpc$
  can be detected above the background.  We show that the emission
  from these regions is substantially greater when we use the
  metallicities predicted by the simulation (which includes enrichment
  through galactic winds) than when we assume a uniform IGM
  metallicity.  Luminous regions correspond to volumes that have
  recently been influenced by galactic winds.  We also show that the
  line emission is clustered on scales $\sim 1 h^{-1} \Mpc$.  We argue
  that although these transitions are not effective tracers of the
  warm-hot intergalactic medium, they do provide a route to study the
  chemical enrichment of the IGM and the physics of galactic winds.
\end{abstract}


\section{Introduction}
\label{intro}

The modern paradigm for cosmological structure formation, the cold
dark matter model, has had admirable success in describing both the
formation of galaxies and the distribution of matter on larger scales.
The model predicts that small perturbations in the early universe grow
through gravitational instability and eventually collapse to form the
objects we see around us.  As collapse proceeds, matter collects in a
``cosmic web'' of mildly overdense sheets and filaments, with galaxies
and galaxy clusters at their intersection.  Simulations and analytic
studies have shown that these gas structures are responsible for the
so-called Ly$\alpha$ forest of absorption lines in quasar spectra
(e.g., \citealt{bi92,bi93,cen94,zhang95,hernquist96,theuns98,dave99}).
However, as observations and simulations have improved, it has become
clear that the intergalactic medium (IGM) has experienced a variety of
complex processes, including shock heating and feedback from galaxies.

Most important, it now appears that a substantial fraction of the
originally pristine intergalactic gas was enriched with a small, but
significant, amount of heavy elements through interactions with
star-forming galaxies.  The best evidence for enrichment comes from
detailed spectroscopic studies of the high-redshift Ly$\alpha$ forest.
Matching individual Ly$\alpha$ absorption features with absorption in
the corresponding wavelength range for metal transitions (especially
the \ion{C}{4} $\lambda\lambda 1548,1551$ and \ion{O}{6}
$\lambda\lambda 1032,1038$ doublets) has revealed that high-column
density absorbers (with $N_{\rm HI} \ga 10^{14.5} \colden$,
corresponding to physical overdensities of a few; e.g.,
\citealt{schaye}) are almost universally enriched to $Z \ga 10^{-3}
Z_\sun$ \citep{tytler,cowie95,songaila96,dave98,carswell02}.
Pixel-by-pixel analyses have revealed the presence of metals in even
lower column density absorbers
\citep{cowie98,ellison00,schaye00,schaye03,aracil03,pieri03}, although
the scatter in the metallicity is large and there is evidence for a
decrease in metallicity with decreasing overdensity
\citep{schaye03}. Metals therefore appear to be widespread in the IGM,
although the mechanism through which they were dispersed remains to be
determined.

The leading candidate is expulsion from star-forming galaxies through
winds driven by supernovae, possibly aided by radiation pressure
(e.g. \citealt{aguirre-dust}).  Such winds have been observed both in
nearby starburst galaxies (e.g., \citealt{lehnert,hlsa,martin}) and in
star-forming Lyman-break galaxies at $z \sim 3$
\citep{pettini01,pettini,shapley03}, but their power, duration, and
filling factor remain unclear.  There is also a theoretical
expectation that similar winds may have occurred in high-redshift
dwarf galaxies, because the small binding energies of these systems
would allow a wind to escape fairly easily
\citep{dekel86,madau,bromm03,furl-metals}.  The time history of
enrichment has important implications not only for the characteristics
of the present-day IGM but also for such topics as the reionization
history of the universe \citep{wyithe,wyithe03,cen, sokasian03} and
ending the era of Population III star formation
\citep{bromm01,oh01,mackey03}.  Detailed studies of the distribution
of metals in the IGM will help to determine how and when enrichment
occurred.  Careful analysis of Ly$\alpha$ forest absorption spectra is
one promising approach.  However, absorption studies of the Ly$\alpha$
forest can only provide information along isolated lines of sight to
quasars, giving no more than an indirect picture of metals in the IGM.
It would thus be useful to find some technique to directly trace the
distribution of metals in three dimensions.

One such method is to search for metal line \emph{emission} from the
IGM.  Because emission does not rely on the existence of a background
source, it can be imaged over large fields of view.  With the addition
of individual line redshifts, we can construct a three-dimensional
picture of the metal distribution (in a similar way to galaxy redshift
surveys).  Observations in this direction have begun: diffuse \osix
emission has been detected from our own Galaxy
\citep{shelton01,dixon01,welsh02} and above the disks of edge-on
spirals \citep{otte03}, while \citet{hoopes03} place limits on the
emission from galactic winds.  Because of the complexity of the
emission processes, and their sensitive dependence on temperature and
density, this question is best addressed through detailed cosmological
simulations.  In this paper, we use such a simulation to calculate
emission from the IGM in the two ultraviolet (UV) doublets \ion{O}{6}
$\lambda \lambda 1032,1038$ and \ion{C}{4} $\lambda \lambda
1548,1551$.  We note that doublets are particularly useful because
they allow an unambiguous identification of the line feature.  We
focus on emission from the low-redshift ($z \la 0.25$) universe that is
most likely to be observable in the near future.  We will contrast a
uniform metallicity with the metal distribution predicted by the
simulation, which includes winds from massive galaxies.

Aside from enrichment, another key process affecting IGM gas is shock
heating: simulations show that a large fraction ($\ga 20 \%$) of
intergalactic gas has $10^5 \kel \la T \la 10^7 \kel$
\citep{cen99,dave01-whim}.  This so-called \emph{warm-hot
intergalactic medium} (or WHIM) is heated through accretion shocks as
sheets and filaments assemble into larger structures.  WHIM gas does
not form a single, well-defined phase: because the shocks are not
uniform, it has a wide range of densities, from near the cosmic mean
to that characteristic of galaxy groups.  But all WHIM gas shares one
crucial characteristic: because it is too warm to form stars but too
cool to produce X-rays, it is very difficult to detect.  The
difficulty is severe enough that gas at or near the cosmic mean
density, whose surface brightness is particularly small, is referred
to as the ``missing baryons.''  One technique to search for WHIM gas
is to use absorption systems \citep{perna98,hellsten98}.
\citet{tripp00} found a large number of low-redshift \osix absorbers
along one line of sight, suggesting that $\ga 10\%$ of baryons are in
the WHIM phase.  However, these absorption studies suffer from the
same drawbacks mentioned above and do not provide a direct picture of
WHIM gas.  Images of its distribution could thus test predictions
about the gravitational growth of structure.  Simulations have shown
that transitions of higher ionization stages of oxygen (\ion{O}{7} and
\ion{O}{8}) are effective probes of the higher-temperature part of
WHIM gas \citep{yoshikawa03}.  We will investigate whether emission
from the \osix and \cfour transitions will allow one to map the
distribution of the lower-temperature portion of the WHIM gas.
Because gas in the appropriate temperature range is expected to lie in
sheets and filaments, this technique would offer the additional
advantage of tracing the cosmic web, providing further confirmation of
the cold dark matter paradigm.

We first calculate the line emissivity of a given gas parcel in \S
\ref{emissivity}.  We then describe our simulation and show how to
calculate the emission from it in \S \ref{sim}.  We discuss our results
in \S \ref{results} and conclude in \S \ref{discussion}.

\section{Metal Emission Mechanisms}
\label{emissivity}

We consider the emission from two metal line doublets: \ion{O}{6}
$\lambda \lambda 1032,1038$ and \ion{C}{4} $\lambda \lambda
1548,1551$.  For each doublet, we use Cloudy 94
\citep{ferland00,cloudy} to compute the emissivity of a given parcel
of gas.  We first construct a grid of densities and temperatures
spanning the entire range of particle properties in our simulation.
We choose density spacings of $\Delta \log (n_{\rm H}/{\rm cm}^{-3}) =
0.25$ for $10^{-6} \leq (n_{\rm H}/{\rm cm}^{-3}) \leq 10^{-2}$ and
$\Delta \log (n_{\rm H}/{\rm cm}^{-3}) = 1.0$ otherwise.  Here $n_{\rm
H}$ is the total density of neutral and ionized hydrogen.  Because the
ionization fractions (and emissivities) are sensitive to variations in
the temperature, we choose $\Delta \log (T/{\rm K}) = 0.05$ for
$10^{4.25} \kel \leq T\leq 10^7 \kel$ and $\Delta \log (T/{\rm K}) =
0.25$ elsewhere.  We then generate grids of the emissivity for each
transition, including collisional or radiative excitation
followed by radiative de-excitation as well as recombination cascades.
However, photon emission following absorption of a photon of the same
wavelength does \emph{not} contribute to the net emission.  We
therefore use Cloudy to compute this ``pumping'' component separately
and subtract it from the total emissivity.  We generate the grids at
solar metallicity; the emissivity is proportional to the
metallicity.\footnote{Actually, the metallicity does affect the
electron density, breaking the strict proportionality.  However, the
difference is negligible for metallicities in the range of interest as
long as the hydrogen is highly ionized (i.e., as long as the gas is
not self-shielded, see below).}  For each of these transitions, the
shorter wavelength line is also the stronger member of the doublet,
with a flux ratio of 2:1.  For concreteness, we quote results for the
stronger line in the following.

\placetable{table:ion}
\begin{deluxetable}{cccccc}
\tablecolumns{6}
\tablewidth{0pc}
\tablecaption{Ionizing Background Models \label{table:ion}} 
  \tablehead{ \colhead{Name} & \colhead{$\log(\Gamma_{\rm HI}/{\rm s}^{-1})$}
    & \colhead{$\Phi_{1026}$} & \colhead{$\Phi_{1216}$} &
    \colhead{$\Phi_{1551}$} & \colhead{$\Phi_{1815}$}}
\startdata 
HM01 & -13.1 & 29.3 & 81.7 & 154 & 182 \\
HM01s & -12.5 & 111 & 309 & 582 & 688 \\
HM01w & -14.0 & 3.5 & 9.8 & 18.4 & 21.8 \\
HM96 & -13.1 & 4.1 & 5.0 & 6.5 & 7.0 \\
Coll & -- & -- & -- & -- & -- \\
\enddata 
\tablecomments{Models labeled `HM96' and
    `HM01' use the spectral shapes of \citet{haardt96} and
    \citet{haardt01}, respectively.  Model `Coll' assumes pure
    collisional ionization.  All surface brightness values are
    reported in $\photfluxA$. }
\end{deluxetable}

Because most of the IGM is photoionized, our choice of ionizing
background could significantly affect the results.  Unfortunately the
ionizing background at $z \sim 0$ is relatively poorly constrained.
\citet{dave01} use observations of the low-redshift Ly$\alpha$ forest
to estimate a total \ion{H}{1} ionizing rate of $\Gamma_{\rm HI} =
10^{-13.3 \pm 0.7} \secinv$ at $z=0.17$.  The shape of the spectrum is
even more poorly known, because the extragalactic background is
difficult to measure directly at these wavelengths, particularly
blueward of Ly$\alpha$.  \citet{brown00} estimate a surface brightness
of $\Phi(1450$--$1900 \mbox{ \AA}) \sim 500 \pm 100 \photfluxA$ based
on \emph{Hubble Space Telescope} observations of diffuse radiation
along a low-extinction line of sight through the Galaxy.  The best
limits in the range 912--1216 \AA\ come from the \emph{Voyager}
spacecraft; \citet{murthy99} place a $1\sigma$ upper limit of
$\Phi(912$--$1216 \mbox{ \AA}) < 30 \photfluxA$, although
\citet{edelstein00} argue that systematic uncertainties increase the
limit to $\Phi(912$--$1216 \mbox{ \AA}) < 10^4 \photfluxA$ at the
$2\sigma$ level.

\begin{figure*}[t]
\plotone{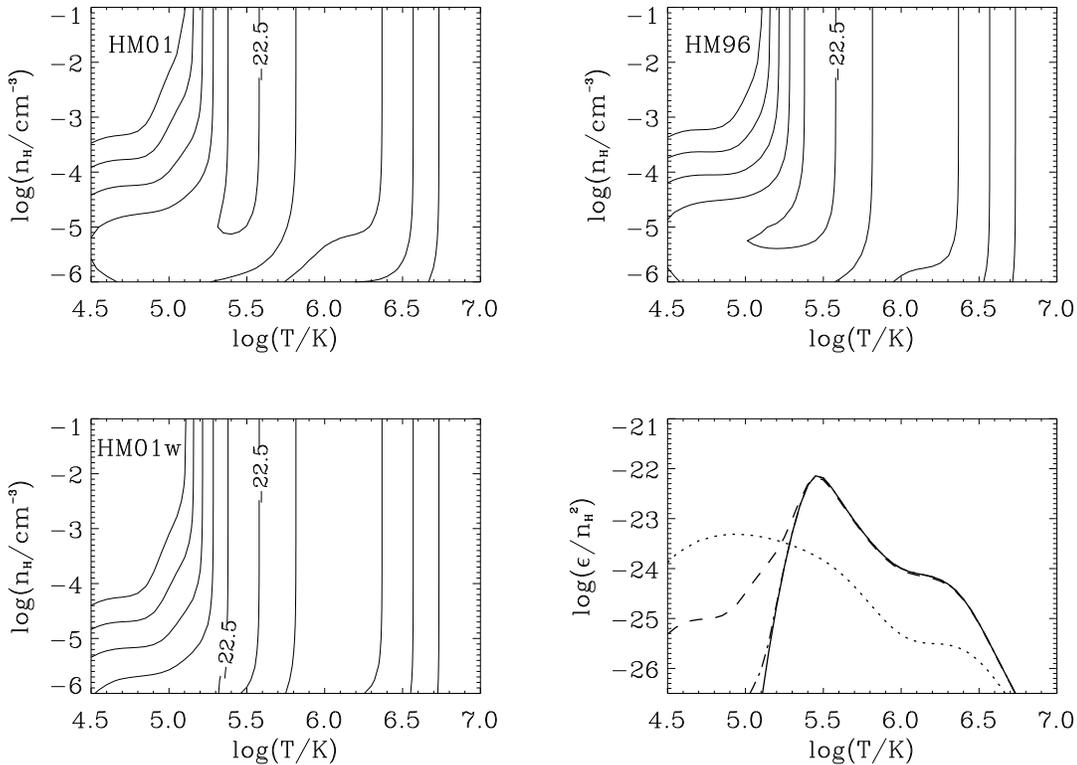}
\caption{ The normalized emissivity $\varepsilon/n_{\rm H}^2$ of
\osix.  All the panels except the bottom right show contour plots in
the $n_{\rm H}$--$T$ plane assuming different ionizing
backgrounds (see text for details).  \emph{Upper left:}
HM01. \emph{Upper right:} HM96.  \emph{Lower left:} HM01w.
\emph{Lower right:} The emissivity assuming collisional
equilibrium (solid curve).  Note that this quantity is independent of
density.  We also show $\varepsilon/n_{\rm H}^2$ for the HM01 spectrum
assuming $n_{\rm H}=10^{-6},\,10^{-4}$, and $10^{-2} \cmden$ (dotted,
dashed, and dot-dashed curves, respectively).\label{fig:o6em}}
\end{figure*}

Because of this uncertainty, we include several choices for the
radiation background.  We base our models on the spectra presented in
\citet[][{\ }hereafter HM96]{haardt96}, which includes only the
contribution of quasars, and \citet[][{\ }hereafter HM01]{haardt01}, which
includes both quasars and star-forming galaxies.  We vary the
normalizations of these spectra over the range allowed by
\citet{dave01}.  We also include a model assuming pure collisional
ionization and excitation.  We summarize the characteristics of our
ionizing backgrounds in Table \ref{table:ion}, including the total
ionizing rate and estimates of the background levels at each of
several wavelengths.  Comparing these results to the direct
measurements described in the previous paragraph, the HM01 model
appears to provide a reasonable fit to the data, although it does
underestimate the flux redward of Ly$\alpha$ reported by
\citet{brown00}.  Note that, even with $\Gamma_{\rm HI}$ fixed, the
HM96 model background is much smaller than that of the HM01 model at
the listed wavelengths.  This is simply because galaxies dominate the
background redward of the Lyman limit, while quasars dominate blueward
of this limit.  Because it does not include galaxies, the HM96 model
strongly underestimates the background at $\sim 1000$--$2000$ \AA.

\begin{figure*}[t]
\plotone{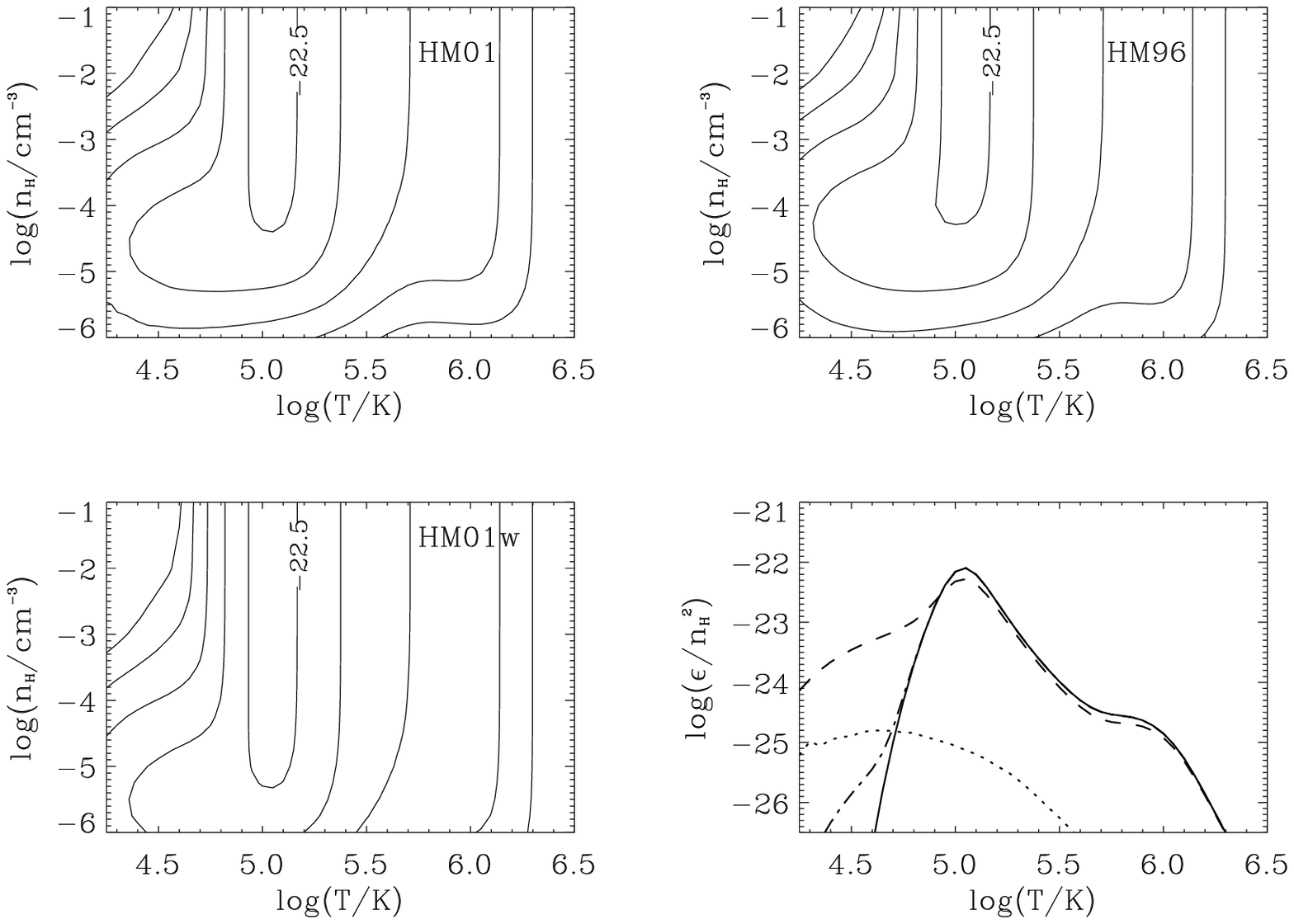}
\caption{ Same as Figure~\ref{fig:o6em}, but for \cfourper.\label{fig:c4em}}
\end{figure*}

The resulting emissivity grids are shown in Figure~\ref{fig:o6em} for
\osix and Figure~\ref{fig:c4em} for \cfourper.  We display, using
logarithmic contours, the density-normalized line emissivity
$\varepsilon/n_{\rm H}^2$ (in units of erg cm$^3$ s$^{-1}$).  In such
units, the emissivity in a purely collisionally ionized gas is
independent of density.  Three of the panels show contour plots of the
emissivity for different choices of the ionizing spectrum: HM01 (top
left), HM96 (top right), and HM01w (bottom left).  We draw contours at
intervals of $\Delta \log(\varepsilon/n_{\rm H}^2) = 1.0$ and label
the highest contour in each panel for clarity.  The bottom right panel
in each figure shows a line plot of $\varepsilon/n_{\rm H}^2$ as a
function of temperature.  The solid line assumes pure collisional
ionization, while the dotted, dashed and dot-dashed curves assume the
HM01 background with $n_{\rm H}=10^{-6},\,10^{-4}$, and $10^{-2}
\cmden$, respectively.  We see that \osix efficiently probes gas with
$T \sim 10^{5.5} \kel$, while \cfour probes gas with $T \sim
10^5\kel$.  Note that the line emissivities are quite sensitive to the
temperature, and a realistic prediction of the IGM emission requires
an accurate treatment of heating and cooling processes.  The bottom
right panels show that photoionization has a substantial effect on the
emissivity of the low-density IGM: it increases the emission from cool
gas but decreases the emission from hot gas.  The photoionizing
background drives low-density gas to higher ionization states than it
would otherwise assume.  On the other hand, these panels also show
that the photoionizing flux makes little difference for gas with
$n_{\rm H}/\bar{n}_{\rm H} \ga 100$.  Such gas is dense enough for
collisional processes to dominate over any reasonable ionizing
background.  A comparison of the HM01 and HM01w panels shows that, for
lower densities, the emissivity quickly returns to the collisional
ionization values as the ionizing photon flux decreases (because the
number of available ionizing photons decreases, so the maximum gas
density that can be affected declines).  Finally, note that the HM96
and HM01 spectra yield nearly identical results, despite the large
differences in the background fluxes for $\lambda \sim 1000$--$2000$
\AA.  This is because we have removed the pumping contribution to the
emissivity and because the ionization potentials for these species are
high, well into the regime in which quasars dominate the spectrum.

One uncertainty in our emissivity calculations is that they implicitly
assume ionization equilibrium.  Testing this assumption precisely is
difficult because equilibrium for high-ionization state metals depends
on a large number of separate recombinations.  However, it appears to
be a reasonable assumption.  We have used Cloudy to compute the
recombination (and collisional ionization) times of \ion{C}{4} and
\ion{O}{6}, along with those of the neighboring ionization states, in
the range $4.75 < \log (T/{\rm K}) < 6.0$.  For metallicities $Z \la
0.1 Z_\sun$, the recombination and ionization timescales are generally
smaller than the cooling timescale by a factor of a few.  For
significantly higher metallicities, the times become comparable.  In
this case, the emissivity of gas near the excitation temperature of a
given transition will decrease (because the relevant ionization state
would be less abundant than expected in equilibrium).  However, the
gas will emit over a wider temperature regime -- and hence for a
longer time -- as it passes through the \ion{O}{6} and \ion{C}{4}
states.  The net effect will be to make the emissivity somewhat less
sensitive to the gas temperature, but only in the highest metallicity
regions.

We also consider two hydrogen lines that may be sources of background
contamination in metal line studies: Ly$\alpha$ $\lambda1216$ (which
may be confused with \cfour emitted from lower redshifts) and
Ly$\beta$ $\lambda 1026$ (which may be confused with \osixper).  If an
instrument can resolve the \ion{O}{6} or \ion{C}{4} doublets, these
lines may be unambiguously identified.  Otherwise, single spectral
lines like the Lyman series will mimic metal emission.  We calculate
the total emission in both Ly$\alpha$ and Ly$\beta$ using Cloudy, with
the same grid spacings as above.  Subtracting the pumping contribution
is complicated by the fact that excitations from absorbed background
photons do not necessarily produce a Ly$\beta$ photon: the excited
state can instead decay to $n=2$ and then to the ground state.  (For
example, a Ly$\beta$ photon can be ``converted'' into Ly$\alpha$ and
H$\alpha$ photons.)  In this case the number of photons removed from
the background radiation field by Ly$\beta$ does not necessarily equal
the number emitted.  We compute the amount of absorption through
\bq 
\varepsilon_{\rm pump} = 2 \pi n_{\rm HI} \hbar \nu B_{12} \bar{J},
\label{eq:lpump}
\eq 
where $n_{\rm HI}$ is the neutral hydrogen density, $B_{12}$ is the
Einstein absorption coefficient, $\nu$ is the frequency of the line,
and $\bar{J}$ is the background flux averaged over the line profile.
For a reasonably smooth background spectrum, $\bar{J}$ is simply equal
to the background flux evaluated at the line center.  We subtract the
amount of energy absorbed from the total emissivity to compute the net
emissivity.  Note that the net result can be \emph{negative} if photon
conversion dominates over collisional excitation and recombination; in
this case we set the emissivity to zero.  We compute and subtract the
pumping contribution to Ly$\alpha$ using Cloudy, just as for the metal
transitions.  We refer the reader to \citet{furl03-lya} for more
information on Ly$\alpha$ emission.

\section{The Cosmological Simulation}
\label{sim}

\subsection{Simulation Parameters}
\label{params}

We perform all of our calculations within the G5 cosmological
simulation of \citet{springel03}.  This smoothed-particle
hydrodynamics (SPH) simulation uses a modified version of the GADGET
code \citep{springel01} incorporating a new conservative formulation
of SPH with the specific entropy as an independent variable
\citep{springel02}.  This ``conservative entropy'' approach
has several advantages over conventional treatments of SPH.  Since the
energy equation is written with the entropy as the independent
thermodynamic variable, the '$pdV$' term is not evaluated explicitly,
reducing noise.  Including terms involving derivatives of the density
with respect to the particle smoothing lengths enables this approach
to conserve entropy (in regions without shocks), even when smoothing
lengths evolve adaptively, avoiding the problems noted by,
e.g., \citet{hernquist93}, and moderates the overcooling problem present
in earlier formulations of SPH.

The simulation assumes a $\Lambda$CDM cosmology with $\Omega_m=0.3$,
$\Omega_\Lambda=0.7$, $\Omega_b=0.04$, $H_0=100 h \hunits$ (with
$h=0.7$), and a scale-invariant primordial power spectrum with index
$n=1$ normalized to $\sigma_8=0.9$ at the present day.  These
parameters are consistent with the most recent cosmological
observations (e.g., \citealt{spergel03}).  The particular simulation
we choose has $324^3$ dark matter particles and (initially) $324^3$
SPH particles in a box with sides of $100 h^{-1}$ comoving Mpc,
yielding particle masses of $2.12 \times 10^9 h^{-1} \msun$ and $3.26
\times 10^8 h^{-1} \msun$ for the dark matter and gas components,
respectively.  The spatial resolution is $8 h^{-1}\kpc$ (comoving).

\begin{figure}[t]
\plotone{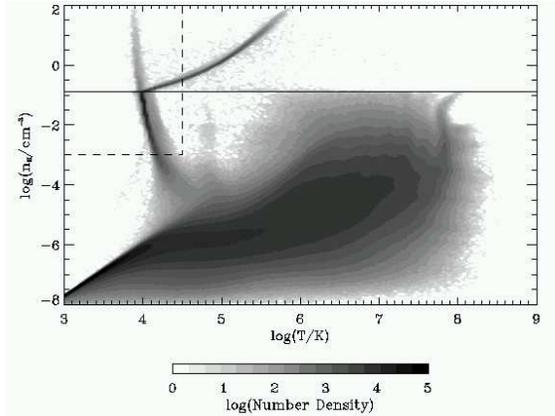}
\caption{ Phase diagram of simulation particles in the box at $z=0$.
The colorscale represents the relative particle density (in
logarithmic units).  The horizontal solid line marks the threshold
density for star formation; we exclude all particles above this line
from our analysis.  Similarly, the dashed box marks particles expected
to be associated with self-shielded regions (also excluded from our
analysis). \label{fig:phase}}
\end{figure}

Figures \ref{fig:o6em} and \ref{fig:c4em} demonstrate that the
emission characteristics of the IGM are extremely sensitive to its
temperature and density.  We show in Figure~\ref{fig:phase} the
distribution of simulation particles (at $z=0$) in the $n_{\rm
H}$--$T$ plane.  Note that the colorscale is logarithmic.  The phase
diagram shows several distinct loci of points.  The majority of
particles lies along a narrow line connecting underdense, cool gas to
moderately overdense, warm gas.  This curve represents the unshocked IGM
gas, and its shape is fixed by a combination of photoheating and
adiabatic expansion \citep{katz96,hui97,schaye99,mcdonald01}.  The
second set of particles occupies a much broader range in temperatures
at moderate to high overdensities. It represents shocked gas in
filaments or virialized halos, including the WHIM gas.  A third set of
particles lies along a nearly vertical line at $T \sim 10^{4} \kel$,
representing gas that has cooled and collapsed into bound objects.
The rapid decrease in the cooling rate at $T<10^4 \kel$ forces such
gas to approach a constant asymptotic temperature slightly below that
value.  Finally, a fourth set of particles lies at high densities
($n_{\rm H} \ga 0.1 \cmden$) and includes all particles able to form
stars.

The simulation includes a new method to describe star formation and
feedback in the interstellar medium (ISM) of galaxies
\citep{springel-sf}.  The model divides the ISM into a cold,
star-forming component and a low density component heated by
supernovae (stars are also included separately).  Gas can move between
the phases through three processes: star formation, the evaporation of
clouds by supernovae, and cooling of the hot phase.  This model yields
a clearly defined density threshold for star formation ($n_{\rm
H}>0.129 \cmden$).  Most important, it yields a converged numerical
prediction for the cosmic star formation rate within the extended
series of simulations performed by \citet{springel03}. In fact, using
simple physical arguments, \citet{hernquist02} have shown that the
evolution of the star formation rate density in these simulations has
a simple analytic form that is determined by the gravitational growth
of halos (the limiting factor at high redshifts, when cooling times
are short) and the expansion rate of the universe (the limiting factor
at low redshifts, when cooling times are long).  We refer the
interested reader to \citet{hernquist02} for more details on the
mechanisms driving the star formation history.  The suite of
simulations performed by \citet{springel03} together with the
converged star formation prediction give us confidence that total star
formation rates in the simulation we have chosen are unaffected by
resolution during the era we study ($z < 0.5$).

The star formation model determines the location of star-forming
particles in Figure~\ref{fig:phase}.  We show the threshold density
for star formation by a horizontal solid line.  For particles above
this threshold, the simulation reports the average temperature of the
cold and hot phases. Dense, actively star-forming gas populates the
narrow locus of particles with mean temperatures reaching 
$\sim10^6\,{\rm K}$, while the faint vertical strip at $10^4\,{\rm K}$
marks particles that have just been selected by the phenomenological
wind model to be ejected from the star-forming phase (see below).
Using the method described in \citet{springel-sf}, we could in
principle separate the cold (star-forming) gas and the hot
(supernova-heated) gas in order to predict the metal line emission
from each phase.  This would be appropriate if we wished to compute
the emission from galaxies themselves.  However, such a treatment
would ignore the inevitable uncertainties in the local ionizing
radiation field (which would be much greater than our assumed diffuse
background), dust content, and the geometry of the two phases within
each galaxy.  Given that the simulation ignores all of these
processes, we chose to simply eliminate all star-forming particles
from our analysis.  As a consequence, our results should be
interpreted as \emph{excluding} emission from galaxies.  Galaxies will
actually emit strongly in the lines we study (at least relative to
pixels without galaxies), but they provide little information about
the IGM and can be ignored for the purposes of this investigation.

We also exclude self-shielded gas from our analysis: by definition,
the ionizing radiation field in such regions is much smaller than the
mean field.  We would therefore overestimate the effects of
photoionization in such regions.  Both simulations \citep{katz96} and
analytic estimates \citep{schaye} show that self-shielding will only
become important in dense, cool regions.  We therefore exclude all gas
with $n_{\rm H} > 10^{-3} \cmden$ and $T < 10^{4.5} \kel$ from our
analysis.  This density threshold marks the point at which the neutral
hydrogen column density $N_{\rm HI} \sim 10^{17} \colden$ (e.g.,
\citealt{schaye}), corresponding to an optical depth to ionizing
radiation of unity.  These criteria are shown by the dashed lines in
Figure~\ref{fig:phase}; we see that self-shielding is only important
for gas on or near the cooling locus described above.  A comparison
with Figures \ref{fig:o6em} and \ref{fig:c4em} show that excluding
self-shielded gas has only a small effect for \cfour and \osixper: the
emissivity of gas below this temperature threshold is so small that
our results are robust to uncertainties in self-shielding.  However,
gas in this region of phase space will produce copious numbers of
Ly$\alpha$ photons.  Our results in this paper therefore underestimate
the number of bright Ly$\alpha$ emitting regions; 
\citet{furl03-lya} consider Ly$\alpha$ emission in much more detail.

Note that we \emph{do} include particles that do not actively form
stars but nevertheless lie near galaxies.  Of course, local sources of
ionizing radiation will also affect such particles.  However, because
bright particles are primarily collisionally ionized (see \S
\ref{uvbkgd}), we do not consider this approximation a bad one.
Figure~\ref{fig:o6em} shows that photoionization becomes important
when the local ionizing radiation field exceeds the diffuse background
by a factor $\sim 100$.  This condition will only be satisfied near
galaxies; including local ionizing sources will probably
\emph{increase} the signal because much of the gas near galaxies is
cooler than the excitation temperature of these transitions.  A more
serious concern is dust: because many of these particles have recently
been ejected from galaxies, they may have significant amounts of dust
that redistributes flux emitted through metal lines.  We do not
attempt to model this process because of the uncertainties involved.

The simulation also includes a model for galactic winds.  Such winds
are the most plausible mechanism for carrying metals from the
star-forming regions in which they are created to the low-density IGM
in which they have now been securely detected (e.g.,
\citealt{ellison00,schaye03}). The volume over which these winds
penetrate is clearly critical for understanding the distribution of
metals (and hence the distribution of \osix and \cfour emission), so
we review the wind model here (see \citealt{springel-sf} for more
details).  Because of the complexity of the wind generation mechanism
in individual galaxies, the treatment must necessarily be
phenomenological.  We first assume that the disk mass-loss rate in the
wind, $\dot{M}_w$, is proportional to the star formation rate:
$\dot{M}_w = \eta \dot{M}_\star$.  Following the observations of
\citet{martin}, we set $\eta = 2$.  We further assume that essentially
all of the available supernova energy powers the wind.  These
conditions fix the wind's initial velocity at $v_w=484 \kms$.  Each
particle in a star-forming region is assigned a probability based on
$\dot{M}_w$ of escaping from the galaxy, and those particles that
escape are given a velocity boost in a random direction in such a way
that momentum is statistically conserved.  Particles that have
recently been selected for ejection appear as a high-density ``tail''
to the cooling IGM gas in Figure~\ref{fig:phase}.  It is worth noting
that we exclude emission from these particles, although in reality
they will contain some enriched, warm gas that may produce metal line
emission.  However, these particles are directly associated with
galaxies and would be difficult to distinguish from their host
galaxies.

Note that the simulation itself included a diffuse ionizing background
with the shape of HM96, normalized as described in \citet{dave99} to
match Ly$\alpha$ forest observations.  This choice leads to
reionization at $z \simeq 6$ and (after this time) largely determines
the temperature distribution of the diffuse gas.  We therefore cannot
include the different heating effects of the ionizing backgrounds
described in Table 1; fortunately this will have only a small effect
on the observable results.  A much more important uncertainty is the
cooling function used in the simulations, which does not include metal
line emission.  To estimate the importance of this effect, we used
Cloudy to compare the cooling times (including metal lines) of gas in
the range $5 < \log(T/{\rm K}) < 6$ for several different
metallicities.  We find that for $Z=0.1 Z_\sun$, the metal lines do
have a substantial effect on the cooling times, decreasing them by a
factor of $2$--$4$ over metal-free gas.  However, the cooling times
fall by only $10$--$30\%$ for $Z=0.01 Z_\sun$.  Because the latter
value is closer to the mean metallicity in the simulation, we do not
expect metal line cooling to have a significant effect in most of the
IGM.  However, it will affect the characteristics of highly enriched
gas, such as that in winds.  We would therefore expect winds to cool
more quickly through the temperature regime in which they emit
strongly, decreasing the number density of such sources by a factor of
a few.  At the same time, each individual source would be somewhat
more compact and hence brighter when it passed through the relevant
temperature range.  However, we expect the real effect of metal line
cooling to be somewhat smaller than this estimate for two reasons.
First, increasing the cooling rates will also increase the amount of
gas able to form stars, which will in turn increase the amount of
energy put into winds.  Second, the above cooling time estimate does
not include cooling due to the expansion of the wind. In semi-analytic
models of winds (e.g. \citealt{madau,furl-metals}), this cooling
mechanism is quite important, so the actual difference between the
cooling times will be smaller than suggested above.

\begin{figure*}[t]
\plotone{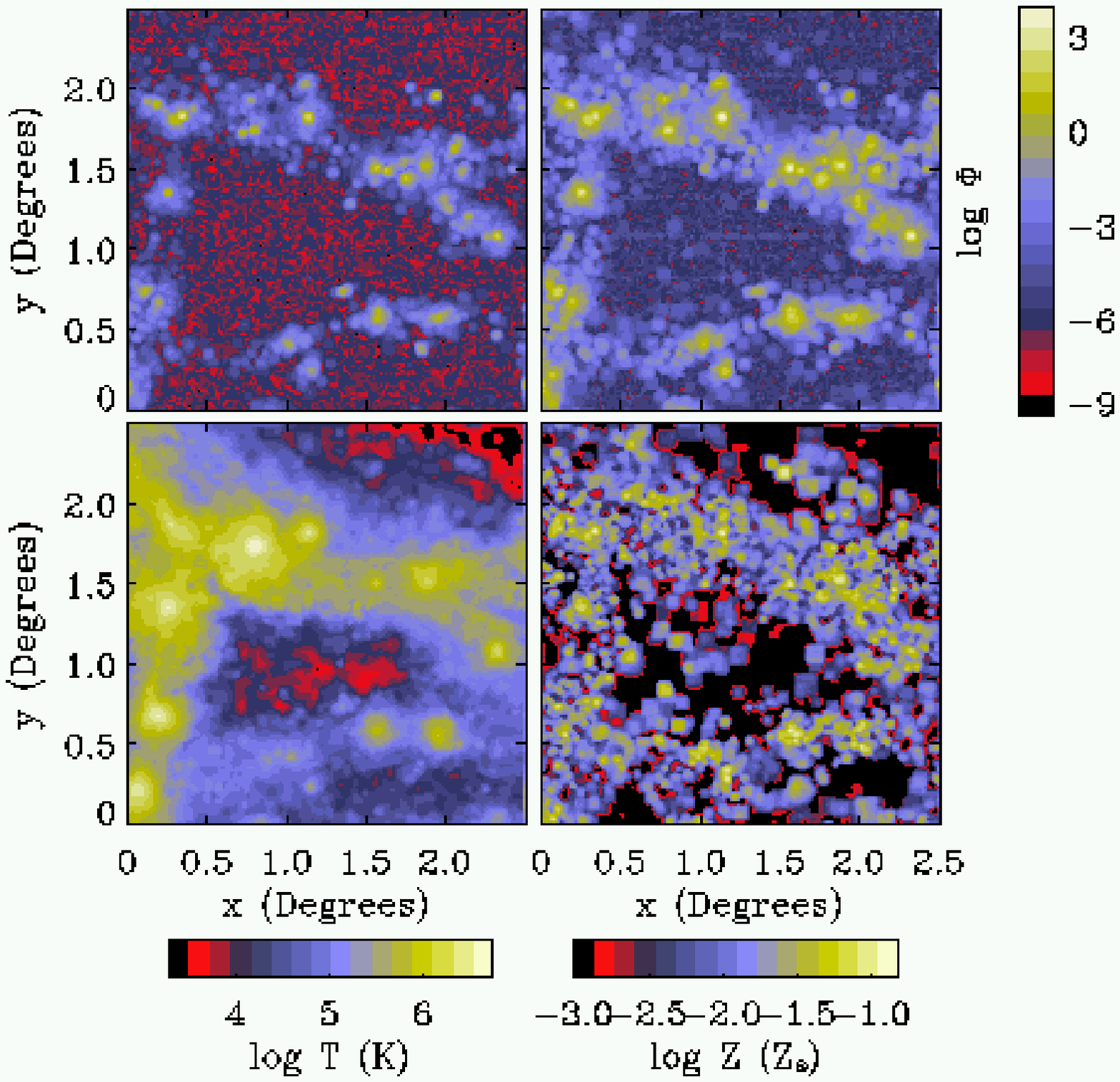}
\caption{ Maps of a $2.5\arcdeg \times 2.5\arcdeg$ slice of the
  universe with $z=0.07$, $\Delta z=0.01$, and $\Delta
  \theta=1\arcmin$.  The top row shows the surface brightness of the
  \cfour (upper left) and \osix (upper right) transitions assuming
  simulation metallicities.  The colorscale is labeled in units of
  $\photflux$.  The bottom row shows the projected mass-averaged
  temperature (lower left) and metallicity (lower right). \label{fig:char}}
\end{figure*}

\subsection{Processing the Simulation}
\label{process}

Given the emissivity grids and the simulation outputs, we now describe
how to calculate the metal line emission from a slice of the universe
at redshift $z$ with a specified depth $\Delta z$ and angular width
$\theta_s$.  We first identify the simulation output (or outputs)
containing this slice.  (Note that the outputs are spaced one
light-crossing time apart, so interpolation between them is
negligible; see \citealt{swh01}).  In order to ensure that we sample a
random volume of the box, we perform the following steps before
processing the output \citep{croft01}.  For each box, we apply a
random translation along each axis, randomly choose an axis of the
simulation box to be the line of sight, and randomly reflect around
each axis.  If $\theta_s < \Theta_{\rm box}/\sqrt{2}$, we also rotate
each particle around the line of sight by a randomly generated angle
(when this criterion is not satisfied such a rotation would interfere
with the periodic boundary conditions imposed on the box).  We then
divide the projected volume into a grid of $N^2$ pixels.  In the
results shown here, we choose $N=600$.

Once these steps are complete, we compute the emission on a
particle-by-particle basis.  For each particle in the box, we first
check whether it lies within the slice of interest.  For this purpose,
we include peculiar velocities in assigning the particle a redshift.
We then assign to each particle within the slice a volume
$V_i=M_i/\rho_i$, where $M_i$ is its mass and $\rho_i$ is its mean
density.  For simplicity, we assume that each particle is cubical with
a uniform density.  In principle, we should distribute the particle
mass and density following the SPH kernel.  However, we find that
luminous particles are typically smaller than our map resolution, so
smoothing would make little difference to our conclusions.
Low-density particles in the diffuse IGM are often larger than our
grid cells, but their emissivity is significantly below any realistic
detection thresholds (see below).  We have checked that the particle
widths given by our procedure are roughly consistent with the SPH
smoothing lengths determined from the simulation.

We then assign a metallicity to each particle.  We use two
prescriptions that contrast two different enrichment processes.  One
option is to use the metallicities reported by the simulation.  These
are ultimately determined by the wind feedback process described in \S
\ref{params} and thus reflect the enrichment pattern expected of
powerful winds from massive galaxies.  Alternatively, we can assume
that metals are distributed uniformly throughout the IGM.  In this
case we normalize the total mass of metals throughout the output box
to that of the simulation.  We note that, in our simulation, the mean
mass-weighted metallicity of the gas at $z=0$ is $Z = 0.024 \, {\rm
  Z}_\sun$.  Of course, for a uniform distribution the emissivity of
each pixel is proportional to the mean metallicity of the simulation
and can be rescaled to any desired level.  We use the solar abundances
of \citet{anders89}: $({\rm C}/{\rm H})_\odot = -3.45$ and $({\rm
  O}/{\rm H})_\odot = -3.13$.

We determine the emissivity of each particle from its density and
temperature by linearly interpolating (in log space) the appropriate
grid described in \S \ref{emissivity} and multiplying by the
metallicity of the particle (in solar units).  The surface brightness
$\Phi_i$ of the $i$th particle (in units of $\photflux$) is then
\bq 
\Phi_i = \frac{1+z_i}{4 \pi r_{L,i}^2} \, \frac{\varepsilon_i}{2 \pi
  \hbar \nu_0} \frac{V_i}{\Delta \Omega_i},
\label{eq:sbptle}
\eq 
where $\nu_0$ is the intrinsic frequency of the line, $r_{L,i}$ is the
luminosity distance to the particle, $\varepsilon_i$ its emissivity,
$z_i$ its redshift, and $\Delta \Omega_i$ the solid angle
subtended by the particle.  We distribute this flux over pixels on the
sky by computing the fraction of the particle volume that overlaps
each pixel.  The total surface brightness of a pixel $j$ is then
\bq
\Phi^{(j)} = \sum_i \Phi_i A^{(j)}_i,
\label{eq:sbpixel}
\eq
where $A^{(j)}_i$ is the fraction of pixel $j$ subtended by particle $i$
and the sum is over all particles.  Finally, we smooth the maps using
a Gaussian filter with a width of four pixels; the resolution $\Delta
\theta$ quoted below is the FWHM of this smoothing function.

\begin{figure*}[t]
\plotone{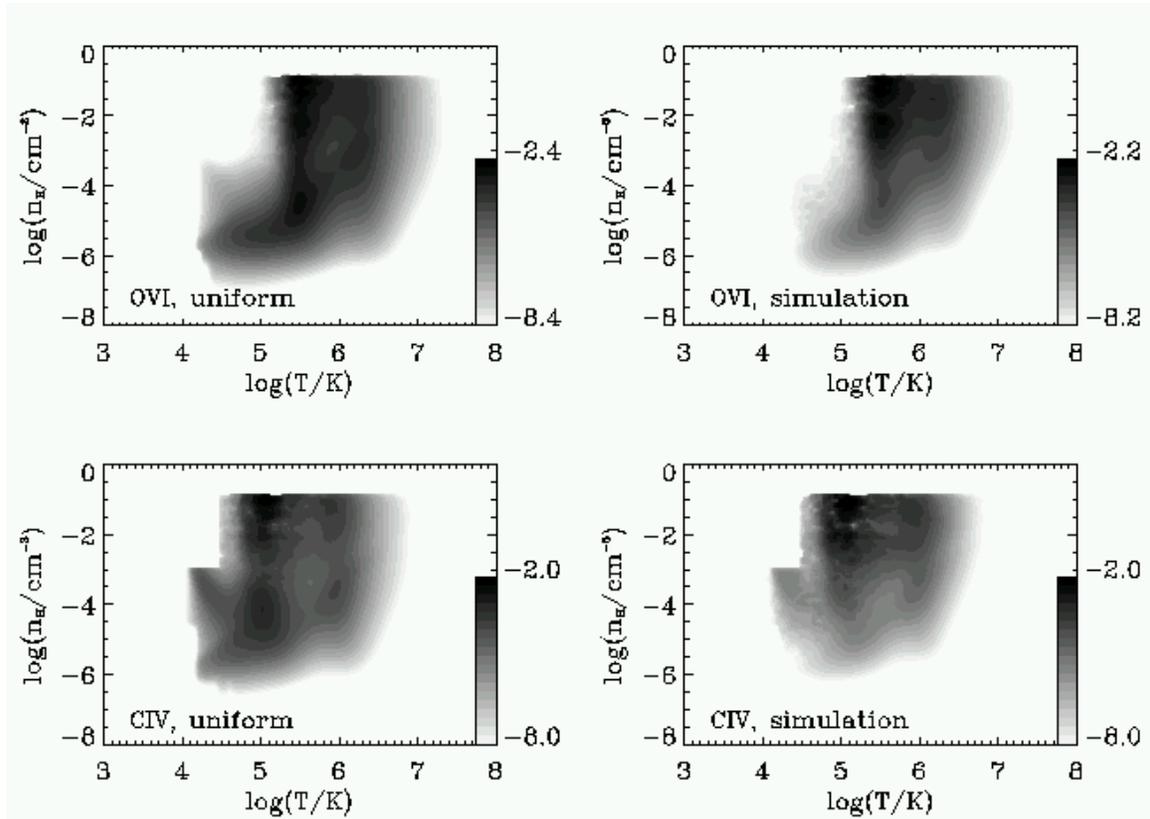}
\caption{ Distribution of simulation particles in the box at $z=0$.
  Each panel shows the relative number density of particles in the
  $n_{\rm H}$--$T$ plane, with a particular weighting.  Note that the
  contours are logarithmic.  The top row weights the number density by
  \osix emissivity for uniform (upper left) and simulation (upper
  right) metallicities.  The bottom row does the same for the \cfour
  transition.  We have excluded star-forming and self-shielded
  particles. \label{fig:phase-wgts}}
\end{figure*}

\vspace{1cm}

\section{Results}
\label{results}

\subsection{Metal Line Maps: Qualitative Features}
\label{maps}

Figure~\ref{fig:char} shows maps of the \cfour and \osix surface
brightness (upper left and upper right, respectively, both assuming
simulation metallicities) together with maps of the projected
mass-averaged temperature (lower left) and metallicity (lower right)
within a slice at $z=0.07$.  The redshift width is $\Delta z=0.01$,
corresponding to $\Delta\lambda \sim 10$ \AA for \osixper; a high
spectral resolution instrument would therefore be able to probe
considerably narrower redshift slices.\footnote{Note that we have
chosen slices that are 10 times thicker (in redshift space) than those
of \citet{furl03-lya}, because the number density of metal emitters is
much smaller than that of Ly$\alpha$ emitters.}  All the maps have
$\Delta \theta = 1\arcmin$ ($\approx 55 h^{-1}$ physical kpc at this
redshift) and use the metallicities reported by the simulation.
Several points are immediately apparent.  First, the two transitions
trace roughly the same structures: it is rare for a region to emit in
one without the other (although this is not necessarily true on a
pixel-by-pixel basis), because they both trace warm high-density gas
(with $T \sim 10^5 \kel$ for \ion{C}{4} and $T \sim 10^{5.5} \kel$ for
\ion{O}{6}).  Filaments are warmer, denser, and more enriched than the
surrounding voids so they stand out clearly in both transitions.  This
is a simple consequence of the metal dispersal mechanism: winds must
begin in galaxies which in turn lie inside overdense regions.  (Note,
however, that the metallicity is quite inhomogeneous even within
filaments.)  Second, the \ion{O}{6} maps better trace the diffuse
filamentary structure of the IGM.  In the \ion{C}{4} maps, filaments
appear to be composed of discrete clumps rather than a smooth gas
distribution; this transition traces cooler gas that is presumably
more tightly associated with galaxies.  Finally -- and most important
-- the mean surface brightness of these transitions is extremely
small, with \osix is significantly stronger than \cfour.  Given the
expected diffuse background ($\ga 10 \photfluxA$ for \osix and $\ga
100 \photfluxA$ for \cfourper; see Table 1), only the brightest
regions will be visible.

Figure~\ref{fig:char} shows that the brightest metal-emitting regions
are associated with hot, approximately spherical regions strung
along filaments.  These regions are heated by two mechanisms.  One is
structure formation: $T \sim 10^6 \kel$ gas is characteristic of large
galaxies and small galactic groups.  The large hotspot in the upper
left of this map is such a group.  A second heat source is galactic
winds.  In our feedback model, the characteristic temperature of the
wind is expected to be $T_{\rm wind} \sim 5 \times 10^6 f_{\rm heat}
\kel$, where $f_{\rm heat}$ is the fraction of the initial kinetic
energy of the wind that remains in thermal energy.  Of course, $f_{\rm
heat}$ is difficult to calibrate because it depends on cooling within
the wind medium (including adiabatic cooling as the wind expands) and
on the gravitational potential from which the wind has to escape, but
ongoing winds can clearly heat gas to the range relevant for these two
transitions.  Winds like the ones in our prescription can reach
distances $\sim 1 h^{-1} \Mpc$ from their host galaxies
\citep{aguirre-wind}, the scale of many of the smaller, nearly
spherical hotspots in Figure~\ref{fig:char}.

\begin{figure*}[t]
\plotone{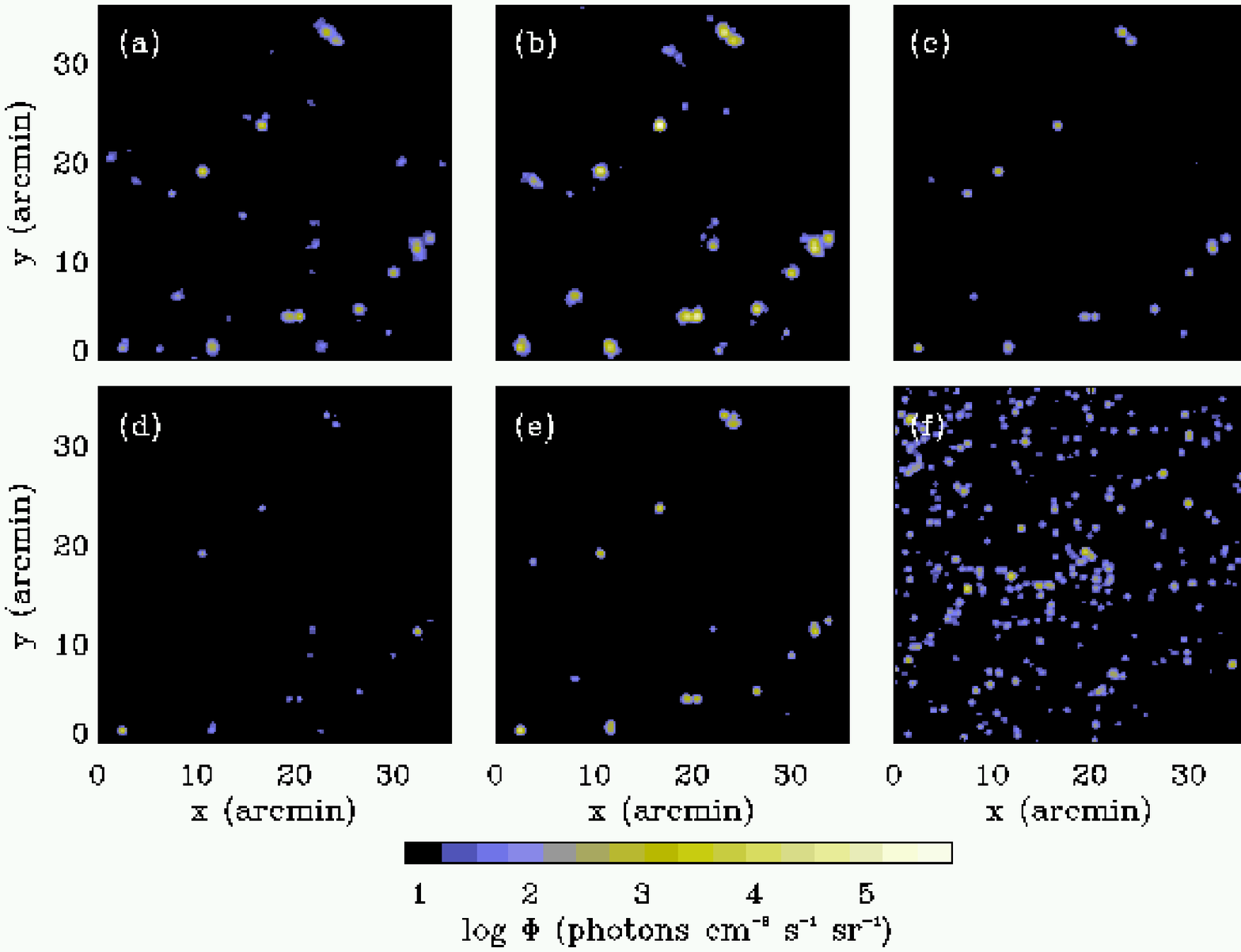}
\caption{ All the panels except \emph{(f)} show maps of a $36\arcmin
  \times 36\arcmin$ slice with $z=0.15$, $\Delta z=0.01$, and $\Delta
  \theta=0.24\arcmin$.  We show only those pixels with $\Phi > 7.5
  \photflux$. \emph{(a):} \osix, uniform metallicity.  \emph{(b):}
  \osix, simulation metallicity.  \emph{(c):} Ly$\beta$.  \emph{(d):}
  \cfour, uniform metallicity.  \emph{(e):} \cfour, simulation
  metallicity. \emph{(f):} Ly$\alpha$, with the same observed
  wavelength range as the \cfour panels. \label{fig:panels}}
\end{figure*}

We can better understand the qualitative features of these maps by
looking more closely at which particles are luminous.  Figure
\ref{fig:phase-wgts} again shows the distribution of particles in the
$n_{\rm H}$--$T$ plane, but this time we have weighted the number
density by the emission in the separate transitions.  The top row
weights the density by \osix emission assuming uniform (left panel)
and simulated (right panel) metallicities.  The bottom row weights the
density by \cfour emissivity.  Note the considerable stretch in the
colorscale; gas in all but the darkest regions will in fact be
unobservable (especially considering that the low-density gas is
distributed over a much larger volume). We see in the figure that, in
all cases, the total emission comes primarily from a small region
within the plane: dense gas in the range $T \sim 10^5$--$10^6 \kel$.
This is reassuring in two ways.  First, a comparison to Figures
\ref{fig:o6em} and \ref{fig:c4em} shows that the gas responsible for
most of the emission is dense enough for its emissivity to be
essentially unaffected by the choice of photoionizing background.  We
thus expect our results to be robust against the uncertainties in the
ionizing background.  Second, note that \osix emission from dense gas
vanishes for $T \la 10^5 \kel$.  Because this is well above our
self-shielding threshold (see Figure~\ref{fig:phase}), the location of
this cut will have little effect on our predictions about this
transition.  \cfour emission does extend below the self-shielding
threshold (the cut is most visible through the horizontal line at
$n_{\rm H} = 10^{-3} \cmden$).  However, most self-shielded gas
rapidly approaches $T \sim 10^4 \kel$ as it becomes denser; such
temperatures are well below the ionization threshold of \cfour and
hence the actual emission from self-shielded gas turns out to be
small.  Furthermore, note that the emission from cooling gas is quite
weak in the model with simulation metallicities, so we would not
expect self-shielded gas to make as much of a difference in this case.

However, the news is not all positive: a comparison to Figure
\ref{fig:phase} shows that most of the emission comes from gas lying
\emph{between} the cooling and IGM loci, especially in the simulation
metallicity model.  The area in which gas emits strongly is only
weakly populated in the simulation and hence metal emission is
concentrated within rare, bright regions.  We also conclude that these
transitions are not efficient probes of gas on the standard IGM
$n_{\rm H}$--$T$ curve or of the shocked WHIM gas.  Finally,
comparison of the two different metallicity models shows that a larger
fraction of emission comes from gas on the IGM and cooling loci if we
assume a uniform metallicity.  This is a simple consequence of the
metal distribution illustrated in Figure~\ref{fig:char}: metals tend
to be more concentrated inside filaments, which are already relatively
bright emitters.  A uniform metallicity thus decreases the contrast
between filaments and voids.

\begin{figure*}[t]
\plotone{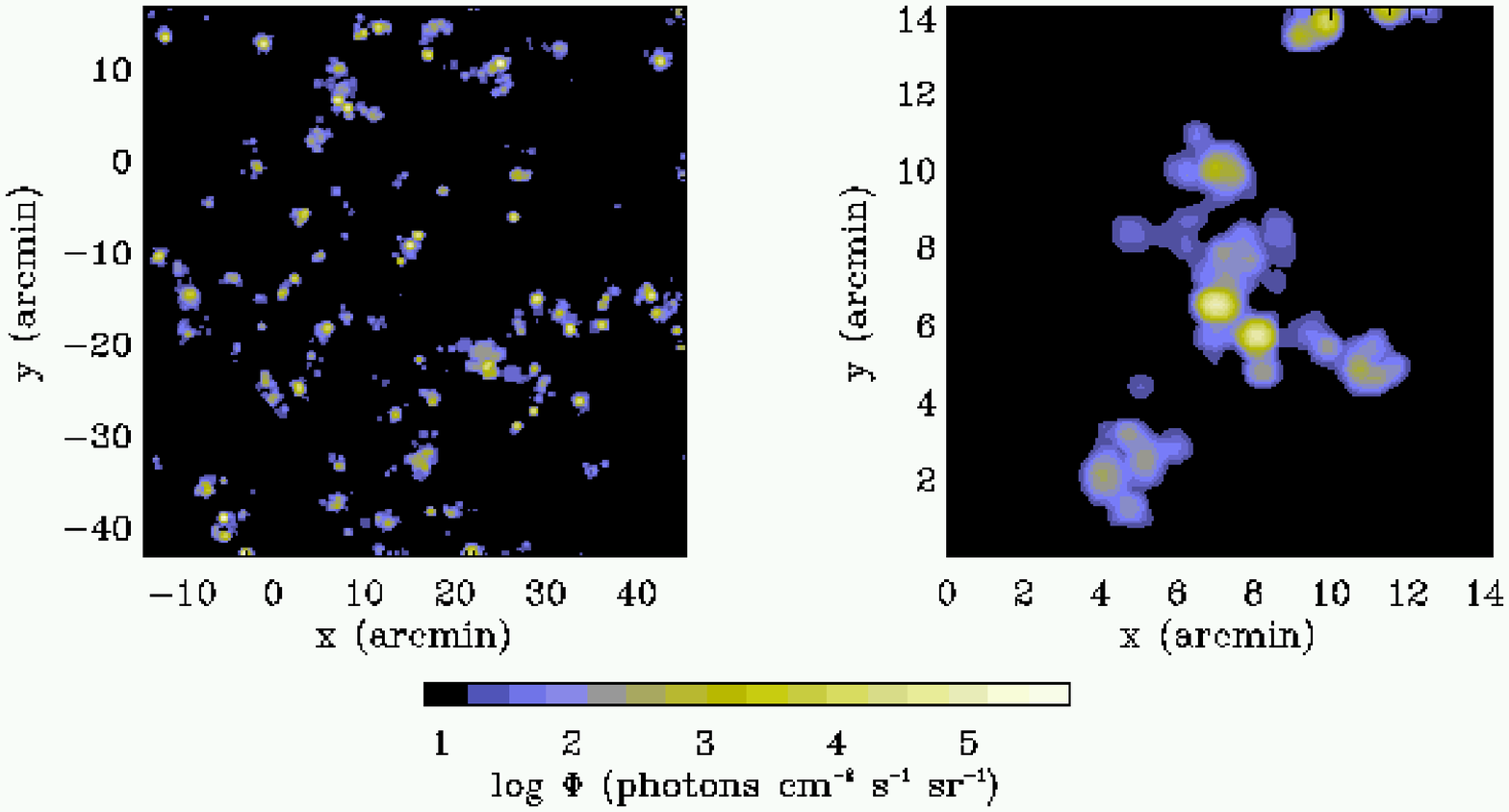}
\caption{ The left panel shows a $1\arcdeg \times 1\arcdeg$ map of the
  \osix transition with $z=0.001$ and $\Delta z = 0.16$, including
  only those pixels with $\Phi > 7.5 \photflux$. The right panel shows
  a closeup of a rich section of the larger map.  Both maps have
  $\Delta \theta=0.24\arcmin$. They are constructed assuming
  simulation metallicities and the HM01 ionizing background.\label{fig:bgcut}}
\end{figure*}

As mentioned above, nearly all of the emission in Figure
\ref{fig:char} is much less than the background and hence
unobservable.  Figure~\ref{fig:panels} shows an example of what one
could see in a mock (background-limited) observation.  Panels
\emph{(a)} through \emph{(e)} show a slice of the universe with
$z=0.15$, $\Delta z=0.01$, and $\Delta \theta = 0.24\arcmin$
(corresponding to a physical size $\approx 25 h^{-1} \kpc$).  Figures
\ref{fig:panels}\emph{a} and \ref{fig:panels}\emph{b} show the surface
brightness of the \osix transition assuming uniform and simulation
metallicities, respectively.  Figures \ref{fig:panels}\emph{d} and
\ref{fig:panels}\emph{e} show the same for the \cfour transition.  We
include only those pixels with $\Phi > 7.5 \photflux$.  Table 1 shows
that this is a reasonable guess for the diffuse background for
\osixper, assuming a spectral resolution $\la 1$ \AA.  It is also
compatible with the upper limit of \citet{murthy99}.  However, it
underestimates, by at least an order of magnitude, the background
redward of Ly$\alpha$ \citep{brown00}.  It will therefore be extremely
difficult to pick out the weaker features shown in the \cfour maps.
In all cases, we see that potentially observable emission is confined
to isolated bright regions.  The number density of such regions is
small, particularly for \cfourper, even though we have chosen a rich
region for illustrative purposes.  This slice has about three times
more bright \osix regions and six times more bright \cfour regions
than average (see \S \ref{pdfs}).  As expected, the peak brightness
increases if we use the simulation metallicities, because these
regions are much more enriched than average.  We find that the typical
sizes of the bright spots are $\sim 80$--$150 h^{-1}$ physical kpc for
\osix and $\sim 40$--$50 h^{-1} \kpc$ for \cfourper.  Clearly, \osix
emission reveals much more interesting structure than \cfour can.

Figure~\ref{fig:bgcut} shows another example of a mock observation.
The left panel shows a $1\arcdeg \times 1\arcdeg$ map of the \osix
transition with $\Delta \theta=0.24\arcmin$ and simulation
metallicities.  We show a full ``light cone'' from $z=0.001$ to
$z=0.16$, illustrating what one could observe with a wide-field
integral field spectrometer, and we again include only those pixels
with $\Phi > 7.5 \photflux$.  The right panel shows a closeup of a
rich section of the larger map.  A typical pointing of such an
instrument would detect a large number of emission regions.  It is
also obvious that bright regions are rarely spherical: many features
have complex substructure that reflects their internal dynamics.

\subsection{Contaminants}
\label{contaminants}

As discussed in \S \ref{emissivity}, emission from neutral hydrogen is
a potential contaminant in searches for these lines.  Figure
\ref{fig:panels}\emph{c} shows the Ly$\beta$ emission from the same
slice as the other panels while Figure~\ref{fig:panels}\emph{f} shows
a different volume corresponding to the same observed wavelength
interval as the \cfour panels (placing it at
$z=0.47$).\footnote{Actually, for a given observed wavelength,
Ly$\beta$ and \osix also trace different volumes.  However, the
wavelength difference between the two transitions is sufficiently
small that we show the same volume for ease of comparison in the
emission mechanisms.  The statistical properties of Ly$\beta$ emission
are unaffected by the small redshift translation.}  Emission from the
Ly$\beta$ line is morphologically similar to that from \cfourper,
though it is even more concentrated.  Strong Ly$\alpha$ or Ly$\beta$
emission requires a relatively neutral, dense medium and so comes
primarily from gas that is near to being self-shielded.  The total
amount of emission in these lines would have been markedly greater if
we had included self-shielded gas, but it would have remained
concentrated in isolated objects (see \citealt{furl03-lya}, which
contains a detailed discussion of Ly$\alpha$ emission).  The
probability of a chance superposition of bright \cfour and Ly$\alpha$
emission remains small.

\subsection{Pixel Probability Distribution Functions}
\label{pdfs}

\begin{figure}[t]
\plotone{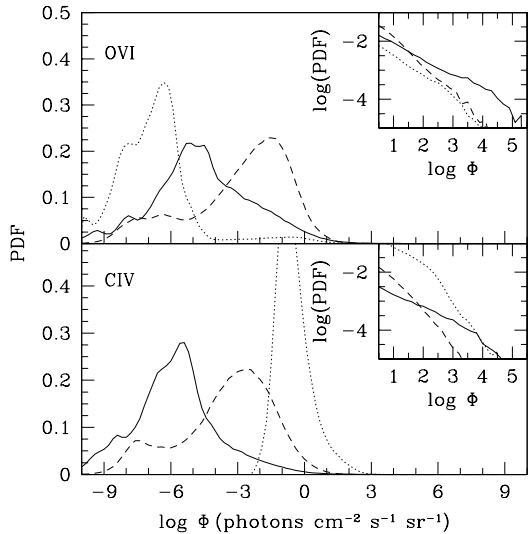}
\caption{ Pixel surface brightness PDFs in our maps for $z=0.15$
(except Ly$\alpha$, see text), $\Delta z=0.01$, and $\Delta
\theta=0.24\arcmin$.  The upper and lower panels show the PDFs for the
\osix and \cfour transitions.  In each panel, the solid and dashed
curves correspond to simulation and uniform metallicities,
respectively.  The dotted curves show the background from Ly$\beta$
and Ly$\alpha$ transitions (upper and lower panels, respectively).
The insets show the bright end of the PDF.  All curves assume the HM01
ionizing background.\label{fig:PDF}}
\end{figure}

Figure~\ref{fig:PDF} shows surface brightness information about our
maps in a more quantitative form.  We have generated an ensemble of
100 maps, each containing $150^2$ pixels with $\Delta \theta =
0.24\arcmin$, and computed the probability distribution function (PDF)
of pixel surface brightness.  Here the PDF is defined as $dn/d \ln
\Phi$ (with unit normalization).  The top and bottom panels show the
PDFs of the \osix and \cfour transitions, respectively.  In each
panel, we show results assuming both simulated (solid lines) and
uniform (dashed lines) metallicities.  The dotted curves show the PDF
for the surface brightness of Ly$\beta$ at $z=0.15$ (top panel) and
Ly$\alpha$ at $z=0.47$ (bottom panel), illustrating the expected
background from these transitions.  The insets show the bright end of
the pixel distribution.  We remind the reader that these histograms
show only the \osix and \cfour components of the doublets; the total
emission from the doublet is $50\%$ greater than the numbers we quote.
On the other hand, if both members of the doublet are required to
confirm the reality of a feature, the weaker member fixes the
experiment's sensitivity.

Bright regions are obviously rare (particularly for \cfourper),
covering only about one percent of the sky in this redshift range.
Because the sources are extended, the PDFs do not translate directly
into space densities.  We find that the comoving density of distinct
sources with $\Phi > 100 \photflux$ is $n_{\rm OVI} \approx 7 \times
10^{-3} h^3 \Mpc^{-3}$ and $n_{\rm CIV} \approx 3 \times 10^{-3} h^3
\Mpc^{-3}$ (assuming simulation metallicities); the difference between
the two is less dramatic than suggested by the PDF because \osix
regions are much more extended.  Clearly, metallicities from the
simulation decrease the surface brightness of the low-density IGM but
dramatically increase the number of bright pixels. Luminous regions
have recently been enriched by galactic winds and hence have above
average metallicity; on the other hand, simulation particles at or
below the mean density of the universe have below-average
metallicities (because they are far from the galaxies that enrich the
IGM).  We also see that PDFs calculated from simulation metallicities
have broad tails extending toward high surface brightness; these
correspond to enriched regions that are moderately overdense and/or
outside of the peak temperature range of the transition (i.e.,
filamentary gas in Figure~\ref{fig:panels}).

Moreover, we find that Ly$\beta$ emission from the low-density IGM is
negligible, even compared to that from \osixper, but that Ly$\alpha$
emission from this gas dominates the metal lines. The difference
between the two hydrogen lines occurs because (1) a much larger
fraction of recombinations result in Ly$\alpha$ photons than Ly$\beta$
and (2) the absorption of a $\lambda=1026 \mbox{ \AA}$ photon by a
neutral hydrogen atom need not result in re-emission of a Ly$\beta$
photon; it can instead be converted to H$\alpha$ and Ly$\alpha$
photons.  In the diffuse IGM, this conversion process dominates over
the Ly$\beta$ photons produced through recombinations and collisional
excitation, so the net emissivity is negligible.  The inset shows that
pixels are more likely to host Ly$\alpha$ emission than \cfour
emission, so having the spectral resolution to resolve both members of
each doublet will be crucial to UV metal line emission studies.
Contamination by Ly$\alpha$ emission becomes even more severe if we
include emission from self-shielded gas \citep{furl03-lya}.

As a check on our methods, we now estimate the mean surface brightness
of the Ly$\alpha$ transition.  We first note that (except in
self-shielded regions) the gas is highly photoionized; in this case
recombinations dominate over other Ly$\alpha$ production processes.
The recombination rate is $\dot{n}_{\rm rec}=\alpha_{\rm H} n_e n_p$,
where $\alpha_{\rm H}$ is the recombination coefficient, and a
fraction $\eta \approx 0.42$ of recombinations produce Ly$\alpha$
photons ($\sim 38\%$ of recombinations go directly to the ground
state, while $\sim 32\%$ of recombinations to excited states pass
through the $^2$S rather than the $^2$P state;
\citealt{osterbrock89}).  The surface brightness per unit wavelength
is then \citep{gould96}
\begin{eqnarray}
\frac{{\rm d} \Phi}{{\rm d} \lambda_{\rm obs}} & = & \frac{\eta
  \dot{n}_{\rm rec}}{4 \pi 
  (1+z)^3} \, \frac{{\rm d} r_{\rm phys}}{{\rm d} z} \, \frac{{\rm d}
  z}{{\rm d} \lambda_{\rm
  obs}} 
\label{eq:lyaestimate} \\
\, & \approx & 3 \times 10^{-3} \left( \frac{\Omega_b
  h^2}{0.02} \right)^2 \frac{(1+z)^2}{h(z)} \nonumber \\ 
& & \times \delta^2 h^{-1}
  \photfluxA, 
\nonumber
\end{eqnarray}
where $h(z) = H(z)/H_0$ and $\delta$ represents the effective
overdensity of the observed pixel.  Given the bandwidth of $\sim 15$
\AA\ in Figure~\ref{fig:PDF}, this simple argument reproduces the peak
of the Ly$\alpha$ PDF with $\delta \sim 1$, giving us confidence in
our analysis procedure.\footnote{One might naively expect a
$(1+z)^{-3}$ dependence from surface brightness conservation (note
that we work in units of $\photfluxA$). However, in reality the
surface brightness increases slowly with redshift because the IGM
density, and hence the recombination rate, increases with redshift.}
We note that this formula only applies to optically thin absorbing
clouds and cannot describe bright pixels in or near the self-shielded
regime \citep{gould96}.  In principle, a generalization of this
formula could be used to estimate the metal line signals as well.
However, the temperature-dependent ionization states of oxygen and
carbon significantly complicate any such attempt; we can really only
extract from it a rough guide to the expected redshift evolution of
metal line emission from the low-density IGM.

\subsection{Angular Resolution}
\label{angres}

\begin{figure}[t]
\plotone{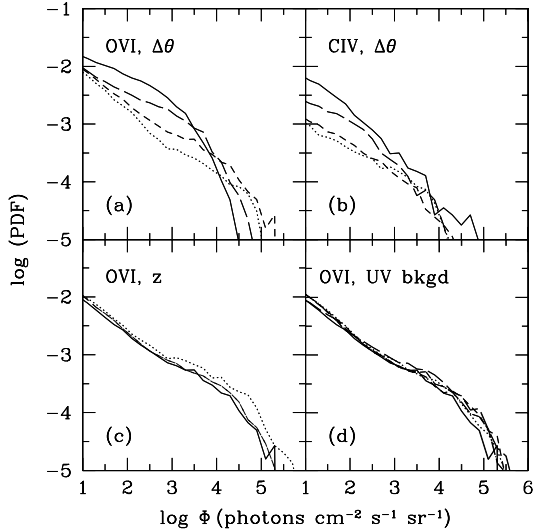}
\caption{ Pixel PDFs.  All curves assume $z=0.15$, $\Delta z=0.01$,
  simulation metallicities, and the HM01 background unless otherwise
  specified. \emph{(a):} The \osix transition with $\Delta \theta =
  1\arcmin,\,0.5\arcmin,\,0.24\arcmin$, and $0.12\arcmin$ (solid,
  long-dashed, short-dashed, and dotted curves, respectively).
  \emph{(b):} Same as \emph{(a)}, but for the \cfour
  transition. \emph{(c):} The \osix transition with $z=0.15,\,0.10$,
  and $0.05$ (solid, dashed, and dotted curves, respectively), with
  fixed physical resolution of $25 h^{-1} \kpc$.  \emph{(d):} PDFs for
  the HM01s, HM01, HM01w, and HM96 ionizing backgrounds (long-dashed,
  solid, short-dashed, and dot-dashed curves, respectively) and for no
  ionizing background (dotted curve), assuming $\Delta \theta =
  0.24\arcmin$.\label{fig:paramhist}}
\end{figure}

As mentioned above, \osix regions have physical scales $\sim 80$--$150
h^{-1} \kpc$ and \cfour regions have sizes $\la 40 h^{-1} \kpc$.
Figures \ref{fig:panels} and \ref{fig:bgcut} show that emitting
regions can have complex substructure.  Typically, \osix emitters have
bright ($\Phi \ga 500 \photflux$) central regions of size $\ga 40
h^{-1} \kpc$ surrounded by irregular, low surface brightness halos.
\cfour regions lack the surrounding halos.  Resolving this
substructure requires reasonably good angular resolution.  We show in
Figures \ref{fig:paramhist}\emph{a} and \ref{fig:paramhist}\emph{b}
the dependence of the bright tail of the pixel PDF on the instrumental
resolution.\footnote{Decreasing $\Delta \theta$ broadens the low
surface brightness peak of the PDF but leaves the mean emission
unaffected.}  Each curve is constructed from 100 maps of $150^2$
pixels each (after smoothing).  The solid, long-dashed, short-dashed,
and dotted curves in each panel have $\Delta \theta =
1\arcmin,\,0.5\arcmin,\,0.24\arcmin$, and $0.12\arcmin$, respectively
(or physical resolutions ranging from $\approx 110h^{-1} \kpc$ to
$\approx 13h^{-1} \kpc$, about twice as large as the simulation
resolution).

We see that the bright tail of the PDF retains qualitatively the same
shape, so our conclusions are essentially independent of angular
resolution.  Increased angular resolution tends to decrease the
fraction of moderately bright pixels, by up to an order of magnitude,
but increase the peak brightness.  The \cfour
distribution changes more with $\Delta \theta$ than the \osix
distribution does because \cfour regions tend to be more compact.
Resolution effects may begin to saturate at $\Delta \theta <
0.24\arcmin$, but this is probably an artifact of being near to the
simulation resolution.

\subsection{Redshift Evolution}
\label{zevol}

The redshift chosen in our Figures is relatively high; \osix has
nearly redshifted to $1216$ \AA, where interference from Ly$\alpha$
will become important. Equation (\ref{eq:lyaestimate}) suggests (and
our numerical results confirm) that the mean surface brightness of the
low-density IGM does not evolve significantly between $z=0$--$0.15$.
However, high surface brightness regions are not well-described by
equation (\ref{eq:lyaestimate}) because they have $\delta > 100$.
This density regime corresponds to regions that have broken off from
the cosmic expansion; such objects should be treated as individual
sources.  Figure~\ref{fig:paramhist}\emph{c} shows the bright end of
the \osix PDF for $z=0.05,\,0.10$, and $0.15$ (dotted, dashed, and
solid curves, respectively).  We have fixed the physical resolution at
each redshift to be $25h^{-1} \kpc$, equivalent to $\Delta
\theta=0.24\arcmin$ at $z=0.15$.  We see very little evolution with
redshift: there is only a slight increase in the number of pixels with
a fixed surface brightness as redshift decreases.  Although an
individual source has a smaller surface brightness as it moves farther
from us, the star formation rate increases rapidly with redshift
\citep{springel03}, so there are more (and more powerful) winds at
higher redshifts.  We find that this effect nearly balances the
increasing cosmic distances.

\subsection{Ultraviolet Background}
\label{uvbkgd}

Figure~\ref{fig:paramhist}\emph{d} shows how the choice of UV
background affects the pixel surface brightness distribution.  The
long-dashed, solid, and short-dashed curves show the PDFs for the
\osix transition assuming the HM01s, HM01, and HM01w backgrounds,
respectively.  The dotted line assumes pure collisional ionization,
while the dot-dashed line assumes the HM96 spectrum.  We see that the
brightest pixels are essentially independent of the choice of ionizing
background.  This is to be expected, because the peak emissivity shown
in Figure~\ref{fig:o6em} comes from collisionally ionized gas whose
location is nearly independent of the background choice.  We note that
the shape of the low-surface brightness peak does depend on the background:
the distribution skews toward higher flux values in weak ionizing
backgrounds, because low-density gas is not photoionized to higher
states (see Figure~\ref{fig:o6em} and discussion thereof).
Although we do not show it, the ionizing background also has no 
effect on bright \cfour emission.

\subsection{Simulation Resolution}
\label{simres}

\begin{figure}[t]
\plotone{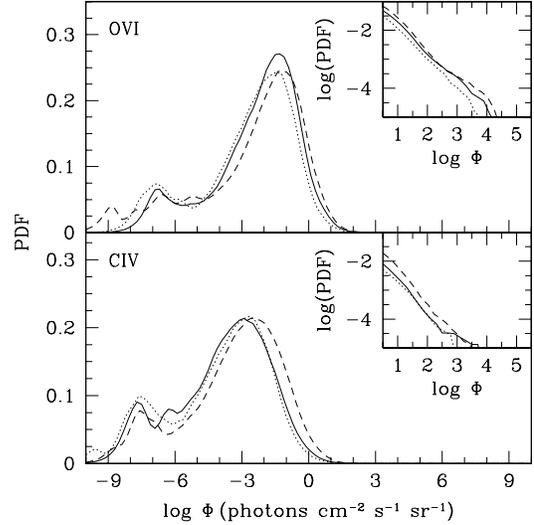}
\caption{ Pixel surface brightness PDFs for 
  $z=0.25$, $\Delta z=0.01$, and a physical resolution of $20 h^{-1}
  \kpc$.  The upper and lower panels show the PDFs for the \osix and
  \cfour transitions.  In each panel, the dotted, solid, and dashed curves
  correspond to the G4, G5, and G6 runs, respectively.  The insets show
  the bright end of the PDF.  All curves assume the HM01 ionizing
  background.\label{fig:simres}}
\end{figure}

One important concern is how the resolution of our simulation affects
the results.  This is particularly important because each bright
region has only of order $\sim 10$ bright particles; we therefore do
not expect their structure to be fully resolved.  Moreover, our wind
prescription is stochastic and will not fully capture the dynamics of
winds in dwarf galaxies.  To estimate the importance of our simulation
resolution, we have repeated our analysis for two other simulations,
G4 and G6, which are identical to the one described above (G5) except
that they contain $144^3$ and $486^3$ particles, respectively (see
\citealt{springel03} and \citealt{nagamine03}).  We show the pixel
PDFs for the different simulations in Figure \ref{fig:simres}; in each
panel, the G4, G5, and G6 simulations are shown with dotted, solid,
and dashed lines.  Both panels assume a uniform metallicity (fixed at
that of the G5 run); we have therefore removed the effects of the
different metal distribution between the simulations.  We see that the
simulation resolution has only a small effect on the resulting PDF: in
particular, the number of bright pixels increases by only a small
amount.  We thus conclude that the temperature and density structure
of the IGM have converged to good accuracy.  The agreement is in fact
surprisingly good, given that the G6 simulation resolves large
galaxies and winds better but also adds many more small galaxies.

The dependence on resolution in the simulated metallicity case is
slightly larger than shown here for the bright pixels, although the
change at small surface brightness is non-negligible between the
three resolutions.  This suggests that the metal distribution does
change between the simulations, probably because of the stochastic
nature of the enrichment process.  If each wind is composed of more
particles, any individual volume element in the simulation is more
likely to contain such a particle, so the metal distribution is less
patchy.  We thus find that the G6 simulation has $\sim 25\%$ more
pixels with $\Phi \sim 10^{-3}$--$1 \photflux$ than the G5 simulation
does.  The patchiness is not as significant for the brightest regions
(where the resolution has about the same effect in both the uniform and
simulated metallicity cases), because such regions are compact and the
``density'' of enriched particles is still large.  Thus the stochastic
nature of the enrichment process only has an important effect at large
distances from the source, where the surface brightness is so low as
to be unobservable anyway.

\subsection{Angular Correlation Functions}
\label{corr}

We define the (flux-weighted) correlation function $w(\theta)$ on a
pixelized map as \citep{croft01}
\bq
w(\theta) = \langle \delta_\Phi(\begin{bf}r\end{bf})
  \delta_\Phi(\begin{bf}{r}\end{bf} + \theta)
\rangle, 
\label{eq:wdefn}
\eq
where $\bf{r}$ is the angular position of a pixel,
$\delta_\Phi(\begin{bf}{r}\end{bf}) =
\Phi(\begin{bf}r\end{bf})/\bar{\Phi} - 1$, and $\bar{\Phi}$ is the
mean pixel surface brightness.  We measure $w(\theta)$ for each pixel
pair of the specified separation and compute the mean value across a
given map.

Because spectral line studies automatically contain redshift
information, angular clustering statistics such as $w(\theta)$ are
perhaps not the optimal measure: one could use the line redshifts to
study clustering in three dimensions.  However, the angular
correlation function is considerably simpler to calculate for a survey
geometry such as ours, and we believe it to be adequate for a first
approach to the problem.  A simple way to retain some redshift
information is to compute $w(\theta)$ only for sources within a
relatively narrow redshift window.  Shot noise will obviously decrease
with bandwidth because more sources are included in the map. However,
using a large bandwidth would sacrifice all redshift information. We
therefore chose to calculate $w(\theta)$ in slices like those
described in the previous section.  This preserves some redshift
information but reduces the shot noise to a manageable level.

\begin{figure}[t]
\plotone{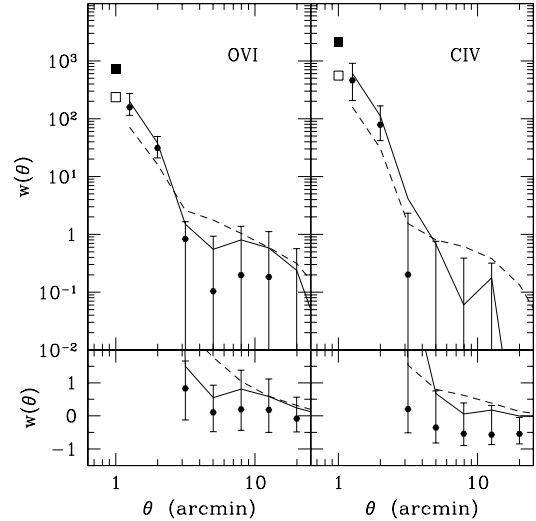}
\caption{ Angular correlation function for the \osix (left) and \cfour
  (right) transitions in slices with $z=0.15$, $\Delta z = 0.01$, and
  $\Delta \theta=1\arcmin$.  The upper and lower panels show
  $w(\theta)$ in logarithmic and linear units, respectively.  The
  solid and dashed curves show the mean signal for the simulation and
  uniform metallicity models, respectively.  The hexagonal points are
  the median values of $w(\theta)$ and the error bars show the upper
  limits of the first and third quartiles (assuming simulation
  metallicities).  The solid and open squares show the autocorrelation
  of the maps for simulation and uniform metallicities, respectively.
\label{fig:angcorr}}
\end{figure}

Figure~\ref{fig:angcorr} shows the angular correlation functions
$w(\theta)$ for the \osix and \cfour transitions (left and right
panels, respectively).  The solid and dashed curves are the mean
correlation function measured from fifty $2.5\arcdeg \times
2.5\arcdeg$ slices with $z=0.15$, $\Delta z=0.01$, and $\Delta
\theta=1\arcmin$, assuming simulation and uniform metallicities,
respectively.\footnote{The typical scale of emitting regions at
$z=0.15$ is $\sim 1\arcmin$, so we are calculating the correlation
function \emph{between} individual systems.}  The solid hexagons are
the median values (assuming simulation metallicities) measured
\emph{between} maps.  The error bars show the upper limits of the
first and third quartiles; they illustrate the dispersion in
measurements of $w(\theta)$ between individual $2.5\arcdeg \times
2.5\arcdeg$ maps.  The solid and open squares show the autocorrelation
of the emission; note that the location along the $\theta$-axis is
arbitrary because $\theta=0$ does not appear on the logarithmic scale.
Because the correlation function is very large for small separations,
the top panels show $w(\theta)$ on a logarithmic scale.  The
correlation rapidly vanishes for larger separations; in order to show
this decrease, we plot $w(\theta)$ on a linear scale in the bottom
panel.

Metal line emission is clearly clustered: $w(\theta=7.5\arcmin) \sim
1$ for \osix and $w(\theta=5\arcmin) \sim 1$ for \cfourper,
corresponding to physical separations of $\sim 800 h^{-1} \kpc$ and
$\sim 550 h^{-1} \kpc$, respectively.  A power law of index $n \sim
3.5$ ($n \sim 4$) adequately fits the correlation function for \osix
(\cfourper).  The correlation function is much steeper than that of
galaxies, at least partly because we have weighted the pixels by their
fluxes.  A difference from galaxies is not surprising because of the
complicated physics behind line emission: there is no reason to expect
the line emission to correspond to bright galaxies.  Rather, emission
is likely to come from those galaxies that (for whatever reason) host
an ongoing wind.  Note that the correlation length we measure is
approximately the size of galactic winds (see Figure~\ref{fig:char}
and discussion thereof).

The case with simulation metallicities is somewhat more strongly
clustered at small scales because of the metallicity gradient between
enriched, emitting regions and the (relatively) pristine, low-density
IGM.  However, differentiating the two through clustering will be
difficult because the dispersion between maps is nearly as large as
the difference between the two models.  Cosmic variance is also very
large, because the correlation function is dominated by rare, bright
pixels.  Comparison of the mean and median shows that a typical map
underestimates the true correlation coefficient, particularly at
intermediate scales.  This is because the positive part of the signal
is dominated by pairs of bright pixels, for which the probability is
quite small.  Thus, a precise measurement of the clustering properties
of metal line emitters will require a survey over a large volume.

We have not included any background emission in this calculation of
$w(\theta)$ because (as discussed above) the background level at these
wavelengths is so uncertain.  However, computing the correlation
function directly from the raw map (without background subtraction)
smooths the correlation substantially because the total flux from a
map of this size is dominated not by the metal emitters but by the
background radiation field.  A better strategy is to set all pixels
below a certain signal-to-noise to zero.  This procedure does not
decrease the number of individual sources by much, because most
contain bright cores well above the background.

\section{Discussion}
\label{discussion}

We have examined the emission expected from the IGM in two UV metal
line doublets, \ion{O}{6} $\lambda \lambda 1032,1038$ and \ion{C}{4}
$\lambda \lambda 1548,1551$.  Using state-of-the-art cosmological
simulations that include prescriptions for star formation and galactic
winds, we computed the surface brightness distribution of these lines.
We assume the IGM to be photoionized by a diffuse background, and we
include several different models for this (currently uncertain)
background radiation.  We also considered two different metallicity
distributions for the IGM: a uniform metallicity model and a model
with the metallicities predicted by our simulation (essentially including
winds from galaxies at low and moderate redshifts).

Can UV metal line emission from the IGM plausibly be observed?  A
comparison of Table 1 and Figure~\ref{fig:PDF} shows that, in the
low-density IGM, the emission from either of our transitions is
several orders of magnitude smaller than that of any realistic
background radiation; we therefore conclude that these transitions
provide little information about the ``missing baryons'' (i.e.,
baryons with high temperatures but low densities).  Instead, Figures
\ref{fig:char} and \ref{fig:panels} show bright emission is
concentrated into isolated regions with (1) significantly higher
metallicity than the average, (2) $T \sim 10^5 \kel$ (for \cfourper)
or $T \sim 10^{5.5} \kel$ (for \osixper), and (3) a density much
greater than the cosmic mean.  \cfour emission is generally confined
to compact regions of size $\la 40 h^{-1} \kpc$, while \osix emission
is generally more extended with bright central cores of size $\sim 50
h^{-1} \kpc$ surrounded by weaker, irregular halos.  The overall space
density of emitting regions is $n_{\rm OVI} \approx 7 \times 10^{-3}
h^3 \Mpc^{-3}$ and $n_{\rm CIV} \approx 3 \times 10^{-3} h^3
\Mpc^{-3}$ (comoving) at $z=0.15$.  Overall, we find that \osix has
more promise as an observational probe.

Detection of metal line emission is feasible (but challenging) with
existing technology.  The ideal instrument for these observations
would be an integral field spectrograph with a large field of view;
resolution of at least $\sim 10\arcsec$ would be necessary to study
the detailed structure of the sources.  Existing instruments do not
meet these criteria, though the next generation of instruments
probably will.  The \emph{Galaxy Evolution Explorer} (\emph{GALEX}),
with a large field of view but relatively low spectral resolution
($\sim 10$ \AA), may detect the brightest cores, though its poor
spectral resolution would make it impossible to identify doublets
unambiguously.  A project underway to construct a balloon-borne
wide-field UV spectrograph will have a limiting sensitivity $\Phi \sim
500 \photflux$ in a single night's observation (D. Schiminovich,
private communication); such an instrument could detect the cores of
many \ion{O}{6} emitting regions.  Finally, the proposed
\emph{Spectroscopy and Photometry of the IGM's Diffuse
Radiation}\footnote{See http://www.bu.edu/spidr/indextoo.html.} would
have the sensitivity to detect the bright cores.  Background-limited
observations to detect emission on $\sim 100 \kpc$ scales will require
a much larger collecting area than currently planned for any
wide-field instrument.  High spectral resolution would help to reduce
the background, which is especially important for observations redward
of $1216$ \AA.

Comparison of the topology of luminous regions to the temperature and
metallicity distribution in the IGM (see Figure~\ref{fig:char}) shows
that most bright particles are embedded in hot ($T \ga 10^6 \kel$)
enriched regions characteristic of galaxy groups and winds. We
therefore suggest that these transitions probe models of metal
enrichment and galactic feedback.  We have shown that these
observations can distinguish a uniform IGM metallicity from one in
which the metals are dispersed by powerful winds.  If mixing is more
efficient than the simulation allows, the emission could be more
extended than we predict but have smaller peak luminosities.
Unfortunately, the feedback parameters in our simulation model are
fixed, and we are thus unable to investigate how the emission varies
with them.  We have deliberately chosen an extreme wind model (in
which none of the supernova energy is lost to radiative cooling before
powering the wind).  Metal emission will tend to be more
compact if winds have less energy simply because they cannot penetrate
as much of the IGM.  However, a more subtle consequence of our choice
is that it allows \emph{all} of the supernova energy to heat the wind
medium.  Weaker winds may therefore lead to \emph{more} emission,
because the wind stays cooler near to the host galaxy; however, note
that the wind velocity of our model is consistent with observations
\citep{martin}.

It is important to note that the extent of observable \osix and
(especially) \cfour emission does not trace out the extent of metal
enrichment.  Collisional excitation of these lines requires high
density and $T \ga 10^5 \kel$.  As winds expand into the IGM, their
density decreases rapidly and thermal energy injected by the wind is
lost to cooling.  Thus we would expect emission from wind-enriched
regions to fade with time.  The declining star formation rate at low
redshifts implies that active winds become relatively rare in the
local universe.  Most enrichment occurred at earlier epochs, giving
the winds time to expand and mix with the surrounding IGM.
Nevertheless, observations of line emission will provide a census of
enriched, warm regions at the present day.

Higher resolution observations of individual sources have the
potential to teach us about the structure of galactic winds.  If
viewed in \osix emission, we predict a central, bright core surrounded
by an irregular, low surface brightness halo.  However, the actual
structure will be much more complex than our simulation, with its
relatively coarse resolution, can accurately describe.  Real galactic
winds have extremely complicated internal structures with several
different temperature phases that can be separated through
observations of diffuse X-ray gas with $k T \ga 0.1 \keV$ (e.g.,
\citealt{strickland}), absorption lines of low-ionization species
(such as \ion{Na}{1}; \citealt{hlsa}), and both emission lines (like
H$\alpha$; \citealt{martin98}) and absorption lines \citep{shapley03}
typical of highly ionized regions.  These are thought to probe,
respectively, a hot phase with large filling factor, a cool phase made
up of embedded clouds, and a warm phase on the interface between the
two \citep{heckman01}.  This last phase should host \osix and \cfour
emission.  Recent observations of the wind surrounding M82, a nearby
starburst galaxy, show that \osix is not an efficient coolant of gas
within several kiloparsecs of the host galaxy \citep{hoopes03};
however, the limiting surface brightness ($\Phi \la 10^5 \photflux$
for this nearby system) is compatible with our predictions.  Because
of the many observational mysteries, metal lines offer a powerful
probe of galactic winds complementary to existing techniques.  In
particular, emission line observations, which can in principle probe
scales up to $\sim 100 \kpc$, will help to determine how winds evolve
on scales large compared to their host galaxies.  Correlating the line
emission maps to galaxy surveys will teach us about the properties of
galaxies that are able to drive powerful winds and by extension the
types of galaxies responsible for enriching the IGM with metals.

Our predictions are subject to several important uncertainties.
First, we have excluded self-shielded and star-forming gas from our
analysis.  Self-shielded gas is unlikely to contribute significantly
to \osix emission, because it is much too cool, but it could add a
small number of additional \cfour sources.  Self-shielding is
significantly more important for studies of hydrogen line transitions
\citep{furl03-lya}, because dense, cool clumps of gas show strong Ly$\alpha$
and Ly$\beta$ lines.  A high enough spectral resolution to resolve the
\ion{O}{6} and \ion{C}{4} doublets will be necessary to prevent
contamination.  Galaxies will inevitably be strong emitters at these
wavelengths and present another potential source of confusion for
studies of the IGM.  However, they can be removed from maps like the
ones we show because they will also be strong continuum emitters,
provided that the angular resolution of the instrument is smaller than
the size of the emitting regions.

Another set of caveats has to do with the simulation itself.  The most
obvious is its finite resolution.  However, we showed in \S
\ref{simres} that the density and temperature structure of our
simulation has converged.  The remaining uncertainty is in the
patchiness of the metal distribution, but that has little effect on
the bright regions.  Another concern is that the simulation neglected
metal line cooling.  We argued in \S \ref{params} that this is
probably only important in highly enriched regions such as winds.  We
expect it to decrease the number density of bright regions by, at
most, a factor of a few, and to increase their surface brightness
(because they cool through the relevant temperature regime sooner).
Note that \citet{hoopes03} argue that line cooling is not important
for winds within $\sim 10 \kpc$ of the host galaxy.

Finally, we neglected local effects such as dust and variations in the
background radiation field.  Because luminous regions tend to be quite
dense, the local radiation field likely will not strongly affect our
results.  Most important, we find that the bright end of the pixel
distribution is largely unaffected by uncertainty in the ionizing
background (see Figure~\ref{fig:paramhist}\emph{d}), although ionizing
radiation from the host galaxy of a wind could still have substantial
effects.  Dust is a particular concern because galactic winds have
been observed to carry it over sizable distances \citep{hlsa}; 
much of our emission comes from regions recently enriched by winds,
so dust will likely decrease the emissivity of such regions.  Fortunately
the measured extinction in local starburst winds is usually not large
and may be patchy, so we expect our qualitative results to remain
unchanged.

Despite these uncertainties, our results show that \cfour and
especially \osix emission can be detected from the IGM.  While
emission from the low density ``missing baryons'' is too weak to be
detected, we find that observations of these transitions may
distinguish different enrichment patterns in the IGM; they thus offer
one of the few methods to probe the effects of mechanical, wind-driven
feedback from galaxies.  Many of the uncertainties we have mentioned
have to do with the detailed structure and mechanism of winds;
observations of the metal emission will of course help to resolve
these questions.  When combined with observations of emission from higher
ionization states (such as \ion{O}{7} and \ion{O}{8};
\citealt{yoshikawa03}), maps of the UV doublets we have studied will
also teach us about the thermal structure of the enriched IGM,
offering an even more complete picture of the wind mechanism.

\acknowledgments

We thank C. Martin, K. Nagamine, and D. Schiminovich for helpful
discussions.  We also thank B. Robertson for assistance with the
computational facilities.  J.S. gratefully acknowledges support from
the W.M. Keck foundation.  This work was supported in part by NSF
grants PHY-0070928, ACI 96-19019, AST 98-02568, AST 99-00877, and AST
00-71019 and NASA ATP grants NAG5-12140 and NAG5-13292.  The
simulations were performed at the Center for Parallel Astrophysical
Computing at the Harvard-Smithsonian Center for Astrophysics.


\end{document}